# Laser operation in non-doped thin films made of a small-molecule organic red-emitter

Hadi Rabbani-Haghighi[1], Sébastien Forget[1], Sébastien Chénais[1], Alain Siove[1], Marie-Claude Castex[1], Elena Ishow[2]

[1]Laboratoire de Physique des Lasers, UMR 7538, Université Paris 13 / CNRS, 93430 Villetaneuse, France
[2]Laboratoire de Photophysique et Photochimie Supramoléculaires et Macromoléculaires,
 UMR-CNRS 8531, ENS Cachan, F-94235 Cachan, France.

Stimulated emission in small-molecule organic films at a high dye concentration is generally hindered by fluorescence quenching, especially in the red region of the spectrum. Here we demonstrate the achievement of high net gains (up to 50 cm$^{-1}$) around 640 nm in thermally evaporated non-doped films of *4-di(4'-tert-butylbiphenyl-4-yl)amino-4'-dicyanovinylbenzene,* which makes this material suitable for green-light pumped single-mode organic lasers with low threshold and superior stability. Lasing effect is demonstrated in a Distributed Bragg Resonator configuration, as well as under the form of random lasing at high pump intensities.

Organic optoelectronics is a very active field of research, partly driven by the rapid developments of organic light emitting diodes and photovoltaic cells. In this context, solid-state organic lasers open the way to many potential applications (in sensing, telecommunications…), due to their ability to produce coherent radiation over the whole visible spectrum, together with low-cost fabrication techniques (spin casting, thermal evaporation) on possibly large areas. While huge challenges remain to be solved in order to achieve direct electrical pumping of organic semiconductors, recent reports have shown the interest of an "indirect" electrical driving strategy, in which a small solid-state laser [1], a laser diode [2] or a light-emitting diode [3] is used as an optical pump source. Stimulated emission and lasing have been observed in numerous conjugated polymers[4-7] as well as in some small molecule-based organic materials[8-10]. The latter are especially attractive as they exhibit a well-defined molecular structure and can be thermally evaporated, which provides an accurate control over the layer thickness and a better film quality. However stimulated emission in small-molecule materials is generally hindered in non-doped films due to strong dipole-dipole coupling between excited-state molecules and π-π stacking interactions (overlap of the π orbitals) between neighboring conjugated segments, which can both lead to significant luminescence quenching[11,12]. To limit self-quenching, the emitter is generally introduced at low concentrations in a polymeric or small-molecule organic matrix[13-16]. Nonetheless, the realization of a single mode (hence very thin) organic waveguide lasers requires high absorption over a short distance, which cannot be readily achieved with low-doped layers. This absorption problem can be bypassed in host-guest blends: in such systems, like the archetypal *tris(8-hydroxylquinoline)aluminium* (Alq$_3$) doped with *4-dicyanomethylene-2-methyl-6-p-dimethylamino-styryl-4H-pyran* (DCM), the host strongly absorbs light and subsequently transfers its excitation to the guest through a Förster energy transfer mechanism[17,18]. In this case the pump radiation is in the ultraviolet, which is detrimental to the photostability of the laser and sets limits in the choice of the pump source.

So far lasing in small-molecule neat films has been demonstrated in materials such as sexiphenyls[19], oligothiophenes[10] or spiro derivatives[20,21]. However, the reported emission wavelength lies essentially in the violet or the blue part of the spectrum, with some rare exceptions in the orange-red[9]. Indeed, long-wavelength emission is generally achieved with extended π-conjugated planar structures, in which the probability to observe π-π stacking (and luminescence quenching) is therefore stronger. In the present work the amplified spontaneous emission (ASE) and laser operation in a Distributed Bragg Reflector-type cavity (DBR) is demonstrated in neat films of an organic dye (*4-di(4'-tert-butylbiphenyl-4-yl)amino-4'-dicyanovinylbenzene*[22], named *fvin* in the following) in the deep red part of the spectrum (650 nm), pumped at 532 nm.

The molecular structure of the *fvin* molecule is depicted in the inset of figure 1. Femtosecond transient absorption spectroscopy studies have evidenced the formation of a quite distorted geometry in the excited state with regard to the ground state. This leads to a large Stokes shift up to 170 nm, with the consequence that red emission is achieved with a relatively short conjugation length, which limits intermolecular π–π aggregation. The large Stokes shift is also favourable to obtain a true four-level system with very low reabsorption losses (see fig. 1). Quenching is also highly reduced here by the steric hindrance exerted by the bulky *tert*-butylphenyl substituents and twisting of the triphenylamino group. These geometrical characteristics allowed us to obtain amorphous materials (T$_g$ = 86 °C) with no microcrystalline areas, which would limit the lasing capabilities because of the induced

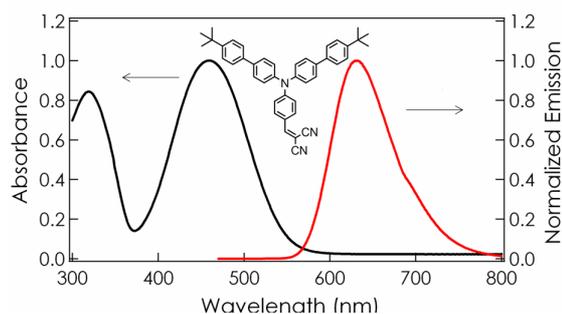

Fig 1: (color online) fvin emission and absorption spectrum in neat film. Inset : structure of the molecule.

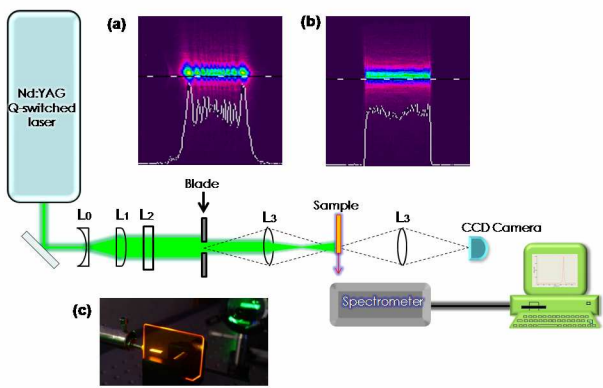

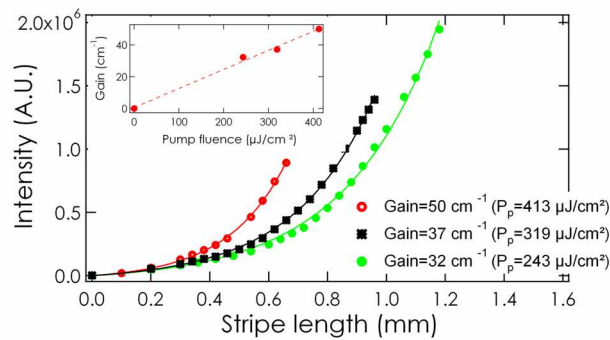

Fig 3 : (color online) net gain (gain minus losses) measurement with the VSL method. Inset : net gain versus pump fluence.

Fig 2 : (color online) experimental setup. Inset : Stripe intensity profile on the sample when the razor blade is set at 10 mm of the sample without imaging lens (A) and with imaging lens (razor blades and sample are conjugates) (B) . C is a photograph of the pumped sample with the collection fiber on the left.

optical losses at the domain boundaries[19].

The absorption and emission spectra of the *fvin* molecule in a neat film configuration is presented in figure 1. The emission band is located in the deep red around 640 nm, whereas the absorption band extends from 400 to 550 nm. The 600-nm thick neat films absorbed as much as 84% of the incident light at 532 nm, in the tail of the absorption band.

The *fvin* neat film was realized on silica substrates through thermal evaporation under a $10^{-6}$ mbar pressure. No visible scattering was observed and the film optical quality was excellent as expected when resorting to the thin film evaporation process. The film thickness was monitored by means of a quartz microbalance in the evaporator machine to be 600 nm. Since the refractive index is 1.82 at 633 nm (determined by ellipsometry), single mode waveguiding is expected.

The variable stripe length method[23] (VSL) was employed to measure the net gain coefficient of *fvin* neat film. In this technique, ASE is detected from the edge of the sample for different pump excitation lengths. The silica substrate was cleaved with a diamond tip to limit scattering. The luminescence was collected via an optical fiber and sent to a spectrometer SPEX 270-M (grating 150 or 1200 lines/mm, corresponding to a resolution of 8 nm or 1 nm, resp.) followed by an Andor technologies DH 720 CCD camera.

The experimental setup is depicted in figure 2. The 532-nm pump beam was produced by a Nd:YVO$_4$ Q-switched laser (PowerChip, Teem Photonics®) with frequency doubling through single-pass in a Lithium Triborate crystal. The pulse duration at 532 nm was 500 ps with a repetition rate of 10 Hz, and the pulse energy could be varied up to 20 μJ. The beam was expanded by an afocal telescope (L$_0$ and L$_1$) and focused with a cylindrical L$_2$ lens to form a thin stripe (320-μm full width at 1/e², measured by a Spiricon camera). After selecting only the central part of this line to provide a uniform excitation intensity onto the sample, an adjustable sharp edge blade, positioned on a translation stage, enabled to vary the length of the stripe. The films were pumped at normal incidence with a pump beam polarization parallel to the sample collection edge. In order to avoid the complex intensity modulation due to Fresnel diffraction effects after the blade edge[24], and as it is not easy in practice to put the blade and the sample very close to each other, the latter were instead optically conjugated through a 1:1 imaging system. We thus managed to produce an intensity profile of the imaged pump stripe with abrupt edges (as shown in the inset of figure 2), provided that the imaging lens had a large enough numerical aperture to collect the high spatial-frequency information contents of the far-field diffraction pattern. In practice, a 10 cm-focal length lens with a diameter of 1 inch was chosen as a good compromise between the desired edge sharpness and the image degradation due to spherical aberration.

Figure 3 shows the ASE intensity versus the pump stripe length for different pump fluences. We clearly observed a superlinear increase of the emission with the excitation length, which can be well fitted with the following expression:

$$I(L) = \frac{\eta_{spont}}{g-\alpha}\left[e^{(g-\alpha)L}-1\right]$$

where I is the ASE intensity, L the stripe length, g is the optical gain (in cm$^{-1}$), α represents the passive losses (self-reabsorption and waveguide losses), η$_{spont}$ is the power density of spontaneous emission emitted into the stripe solid angle.

In this expression, the gain saturation is neglected, which is reasonable as only short pump stripes are considered. The amount of losses α were experimentally measured with a similar method[23] and were estimated to be around 10 cm$^{-1}$. We obtained a maximum optical gain of 50 cm$^{-1}$ for a pump fluence of 413 μJ/cm², which ranges among the highest reported for red emitting neat films of small molecules. The ASE spectrum is shown in the insert of the figure 3. The maximum emission wavelength is in the deep red part of the spectrum, around 660 nm and the ASE threshold is 22μJ/cm², which is the lowest reported ASE "threshold" (as defined by Mc Gehee[25]) for neat films made of small-molecule organic red-emitters. Moreover, pumping in the green instead of UV induces excellent ASE stability: 80% of initial ASE intensity (just above threshold) was still measured after $10^5$ pulses at 10 Hz.

For the sake of comparison, we used the well known DCM dye in the same single mode waveguide configuration since it exhibits similar absorption and emission spectral characteristics. Contrarily to *fvin*, DCM is very sensitive to quenching, which limited us to a 5 wt. % doping in PMMA matrix (Microchem Inc, M$_w$ = 4.95×10$^5$ g mol$^{-1}$). The sample was prepared by spin coating the DCM-PMMA anisole solutions on Silicon:SOG substrates at a spin speed of 1000 rpm to obtain a 600-nm thick film. In this case only 15% of the pump beam at 532 nm was absorbed. With the VSL method, an optical gain of 40 cm$^{-1}$ was measured at the maximum pump fluence (413 μJ/cm²).

For a similar absorbed pump power, the gain in DCM is then much higher than in *fvin*. However, as the film thickness was fixed to ensure single mode waveguiding, it is noticeable that a higher gain is obtained with *fvin* for the same incident pump power.

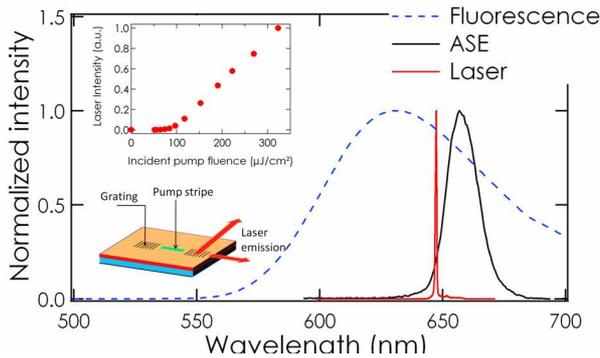

Fig.4 : (color online) fluorescence, ASE and laser spectrum in the DBR configuration. Inset : Laser chip design (here the pump stripe length is 2 mm).

The *fvin* capabilities in a laser cavity were first tested upon appending a DBR resonator to the waveguide. To obtain a periodic modulation of the film thickness, we used a very simple all-optical method based on direct laser ablation through a phase mask. The experimental setup is described in detail in ref. 26. We used an ArF excimer laser ($\lambda$=193 nm, energy per pulse = 300 μJ, repetition rate of 10 Hz) to illuminate a phase mask: the diffracted orders interfere in the vicinity of the mask, creating an interference pattern with a pitch that can be shown to be, under the conditions of this work, the same as the period of the mask[26]. Local ablation of the organic material is obtained for illumination times on the order of a minute (several hundreds of pulses).

The optimal pitch $\Lambda$ for a Bragg grating is given by the Bragg formula $2.n_{eff}.\Lambda = m\lambda$, where $n_{eff}$ is the effective index of the waveguide and m is the Bragg order. Here, we used a mask with a 1090 nm period, leading to laser emission at the m=6 order. The periodicity of the grating extends over several millimeters with a modulation depth of 200 nm. The grating structure appears under an AFM microscope to be smooth, free of redeposition debris and groove defects.

For the sake of demonstrating the existence of a laser effect, this unoptimized phase mask was chosen because it was readily available. However the laser threshold could be lowered by using a lower-order resonator made by a lower-pitch phase mask.

We used a DBR configuration in which two Bragg gratings (with the same 1090 nm pitch) facing each other were engraved in a PMMA layer before the evaporation of a 695 nm thick *fvin* layer (see inset in figure 4). The pump stripe (active region) was located between the two mirrors. Several samples with various distances between the two Bragg mirrors were realized. We observed laser operation as demonstrated in figure 4. The laser spectrum peaks at 647 nm as expected from the Bragg formula with $n_{eff} = 1.82$. The polarization of the laser beam was parallel to the layer (TE). A typical curve showing the variation of the laser intensity with pump power is shown on figure 4 (inset). The lasing threshold is relatively high (60 μJ/cm²) because of the high order Bragg configuration.

Interestingly, we also observed that under high pump fluences, lasing can occur without any resonator, under the form of *random lasing*, a phenomenon that has been already observed in some films of conjugated polymer and small molecules[27]. Indeed, we observed that the ASE spectrum, for fluences higher than typically 0.3 mJ/cm², was highly structured, exhibiting multiple narrow peaks with a spectral width limited here by the spectrometer resolution (1 nm). The physical origin of the feedback still remains unclear in this system and is under study.

In summary, we have demonstrated single mode laser action in a thermally evaporated neat film of a red-emitting dye at 647 nm, with a very simple DBR patterning technique. The optical gain was measured for different pump powers with the variable stripe length method and a maximum value of 50 cm⁻¹ was obtained. In order to obtain more accurate fits, a simple way to avoid the complex intensity modulation provided by Fresnel diffraction on the edges of the slit was presented. The fact that *fvin* molecule exhibits optical gain and ASE with a threshold as low as 22 μJ/cm² in a neat film configuration shows that quenching effects are tremendously reduced compared to classical dyes such as DCM or rhodamine, in which emission is totally suppressed in neat films. Consequently, in a single mode waveguide configuration, a higher gain can be obtained for the same pump power in a neat film of *fvin* than in a doped film of DCM. Laser operation in the red (at 647 nm) was demonstrated with a DBR resonator fabricated with a very simple technique. At last, random lasing with coherent feedback was observed for pump fluences exceeding ~0.3 mJ/cm².

The present results give very favourable implication for the use of *fvin* material as red-emitting organic laser material with a pumping wavelength in blue or green region.

We acknowledge the ANR ("Bachelor" project) and the Université Paris 13 (BQR credits) for funding this work. We thank M. Lebental for loaning us the DCM sample used for these experiments. We thank P. Sebbah for fruitful discussions on random lasers and D. Ades for assistance in making the PMMA films.